\documentclass[prb,aps,epsfig,twocolumn,showpacs]{revtex4}

\usepackage{epsfig,amssymb,amsmath,latexsym}

\newcommand\beq{\begin{equation}}
\newcommand\eeq{\end{equation}}

\newcommand\bea{\begin{eqnarray}}
\newcommand\eea{\end{eqnarray}}

\newcommand\bi{\begin{itemize}}
\newcommand\ei{\end{itemize}}

\newcommand\non{\nonumber}

\newcommand\up{\uparrow}
\newcommand\dn{\downarrow}

\newcommand\smat{$\mathbb S$}

\newcommand\hmi{{\mathcal H}_{\textsf{int}}}

\newcommand\epp{e^{i\,(k\,+\,k_F)\,x}}
\newcommand\epm{e^{i\,(-k\,+\,k_F)\,x}}

\newcommand\emp{e^{-i\,(k\,+\,k_F)\,x}}
\newcommand\emm{e^{-i\,(-k\,+\,k_F)\,x}}

\newcommand\odd{1{\textendash}D}

\newcommand\od{1{\textendash}D~}

\newcommand\afp{{\textsf{AFP~}}}

\newcommand\afpd{{\textsf{AFP}}}

\newcommand\cafp{{\textsf{CAFP~}}}

\newcommand\cafpd{{\textsf{CAFP}}}

\newcommand\rfp{{\textsf{RFP~}}}

\newcommand\rfpd{{\textsf{RFP}}}

\newcommand\tfp{{\textsf{TFP~}}}

\newcommand\tfpd{{\textsf{TFP}}}

\newcommand\sfp{{\textsf{SFP~}}}

\newcommand\sfpd{{\textsf{SFP}}}

\newcommand\cfpd{{\textsf{CFP}}}

\newcommand\cfp{{\textsf{CFP~}}}

\newcommand\gfpd{{\textsf{GFP}}}
\newcommand\gfp{{\textsf{GFP~}}}

\newcommand\lld{{\textsf{LL}}}
\renewcommand\ll{{\textsf{LL~}}}

\newcommand\ctd{{\textsf{CT}}}

\newcommand\ct{{\textsf{CT~}}}

\newcommand\wirgd{{\textsf{WIRG}}}

\newcommand\qwd{{\textsf{QW}}}

\newcommand\nsnd{{\textsf{NSN}}}

\newcommand\card{{\textsf{CAR}}}
\newcommand\ard{{\textsf{AR}}}

\newcommand\ed{{\textsf{e}}}

\newcommand\ns{{\textsf{NS~}}}
\newcommand\rg{{\textsf{RG~}}}
\newcommand\qw{{\textsf{QW~}}}

\newcommand\nsn{{\textsf{NSN~}}}

\newcommand\car{{\textsf{CAR~}}}
\newcommand\ar{{\textsf{AR~}}}

\newcommand\butt{\rm{B\"uttiker}~}

\def\dfrac#1#2{{\displaystyle\frac{#1}{#2}}}

\newif\ifboo \boofalse

\begin{document}

\textheight=23.8cm

\title{\Large
A systematic stability analysis of the renormalisation group flow
for the normal-superconductor-normal junction of Luttinger liquid
wires}

\author{\textsf {Sourin Das$^{1}$, Sumathi Rao$^2$ and Arijit Saha$^2$}}
\affiliation{ $^1$ {Centre for High Energy Physics, Indian
Institute of Science, Bangalore 560 012, India}} \affiliation{ $^2$
 {Harish-Chandra Research Institute, Chhatnag Road, Jhusi,
Allahabad 211019, India}}

\date{\today}


\begin{abstract}
We study the renormalization group flows of the two terminal
conductance of a superconducting junction of two Luttinger liquid
wires. We compute the power laws associated with the renormalization
group flow around the various fixed points of this system
using the generators of the $SU(4)$ group to generate the appropriate
parameterization of an \smat-matrix representing small deviations
from a given fixed point \smat-matrix (obtained earlier in Phys. Rev. {\bf B
77}, 155418 (2008)), and we then
perform a comprehensive stability analysis.
 In particular, for the  non-trivial fixed point which has intermediate
values of transmission, reflection, Andreev reflection and crossed
Andreev reflection, we show that there are eleven independent
directions in which the system can be perturbed,
which are relevant or irrelevant, and five directions
which are marginal. We obtain power laws associated with these
relevant and irrelevant perturbations. Unlike the case of the two-wire
charge-conserving junction, here  we show that there are power
laws which are non-linear functions of $V(0)$ and $V(2k_F)$ (where
$V(k)$ represents the Fourier transform of the inter-electron
interaction potential at momentum $k$). We also obtain the power law
dependence of linear response conductance on voltage bias or temperature
around this fixed point.
\end{abstract}

\maketitle

Electron-electron ($\ed$-$\ed$) interactions in low-dimensional
systems  (one-dimensional (\odd) quantum wires (\qwd) and dots) can
lead to non-trivial low energy transport properties due to the
Luttinger liquid (\lld) ground state of the system. In this context,
a geometry which has gained considerable attention in the recent
past is the multiple \ll wire junction. In general, junctions of
multiple \qw can be viewed as quantum impurities in a \ll from which
electrons get scattered at the junction. For the simplest case of
two-wires, the junction can be modeled as a back-scatterer while for
the general case of multiple \qwd, the junction represents a more
non-trivial quantum impurity which may not be as straightforward to
model microscopically.

For the two-wire system, it is well-known~\cite{k&f,yue} that in the
presence of a scatterer, there are only two low energy fixed points -
{\sl {(i) the disconnected fixed point}} with no transmission
(i.e. the transmission amplitude for incident electron or hole,
$t=0$) which is stable and  {\sl {(ii) the transmitting fixed
point}} with no reflection ($t=1$) which is unstable. More recently,
the low energy dynamics of multiple \ll wires connected to a
junction have also been studied in
detail~\cite{nayak,lal,rgstudy_drs,affleck1,affleck2,bellazzini2006,
bellazzini12008,bellazzini22008} and several interesting fixed points have
been found, including continuous one-parameter families of fixed
points~\cite{tdos_adrs}. These studies have also been
generalized
theoretically~\cite{jap1,jap3,maslov1996PRB,fazio,bena,man,dloss,titov,
winkelholz} to describe a junction of \od  wires with superconductors and
have also been generalized to include spin~\cite{chamon2008}.
Our recent work~\cite{nsnepl_drs,nsnprb_drs,superduality_dr} has
generalized these studies to the case of superconducting junction of
multiple \qwd. In such a system, due to the proximity of the
superconductor, both  electrons as well as holes take part in the
transport which leads to very interesting transport properties at
small bias resulting from the interplay of \ll correlations and the
proximity induced pair potential.

In this article, we study the case  where  two \ll \qw are
coupled simultaneously to a bulk superconductor. The physical
separation between the junctions of the two-wires with the
superconductor is of the order of the size of the Cooper-pair. This
leads to the realization of a normal-superconductor-normal (\nsnd)
junction which allows for direct tunneling of electrons from one wire to
the other and also allows a finite amplitude for the crossed Andreev
reflection (\card)
process~\cite{hekking1,hekking2,russo,chandrasekhar,yeyati,zaikin2007,
zaikin12007,konig2009,belzig2008,belzig2006,zaikin2009} in addition to
the normal reflection and Andreev reflection (\ard) processes.

In an earlier study of the \nsn junction~\cite{nsnprb_drs}, we showed
that the \nsn junction has more than two fixed points unlike the
normal two-wire junction (as mentioned above) or the junction of \ll with
a bulk superconductor (\ns junction) which has only two fixed points  -
{\sl {(i) the Andreev fixed point}} where the amplitude for Andreev
reflection (\ard), $r_{A}=1$ and normal reflection amplitude, $r=0$ and
which is unstable and {\sl {(ii) the disconnected fixed point}} where
$r_A=0$ and $r=1$, and which is stable~\cite{jap1,jap3}. We showed that
there exists a fixed point with intermediate values of transmission and
reflection. Thus, the \nsn junction is the minimum configuration which
possesses non-trivial fixed points with intermediate transmission and
reflection amplitudes. In what follows, we will focus  mainly on the
\nsn junction.

In the previous studies, a comprehensive analysis of the various
possible perturbations allowed by symmetry around all the fixed
points of the \nsn junction was lacking. In this article, we carry
out a systematic stability analysis for each of the fixed points
obtained earlier in Ref.~\onlinecite{nsnprb_drs} for the \nsn
junction and we predict the power laws associated with all possible
independent perturbations that can be switched on around these fixed
points. Our analysis provides us with renormalized values of the
various transmission and reflection amplitudes around these  fixed
point values which can then be used to obtain the Landauer-\butt
conductances.

We start with a brief review of the \rg method followed in
Refs.~\onlinecite{yue} and \onlinecite{lal} where an \smat-matrix
formulation was used to compute the linear conductance and
inter-electron interactions inside the \qw were taken into account by
allowing the \smat-matrix to flow as a function of the relevant
energy scale (like temperature, bias voltage or system size) using an
\rg procedure. This method works well when $\ed$-$\ed$ interaction
strength inside the \qw is weak so that it can be treated
perturbatively. This is usually referred to as the weak interaction
renormalization group (\wirgd) procedure~\cite{matveev}.

To benchmark our calculation with  known results, we first
calculate the complete set of all possible power laws associated
with the independent perturbations that can be switched on around
the time-reversal symmetry broken chiral fixed points (\cfpd) and
the time reversal symmetric Griffiths fixed point (\gfpd) of a
normal junction of three \ll wires~\cite{lal}. Then we apply the
same procedure to the case of the \nsn junction. The strength of our
formulation to obtain the power law scaling of the perturbations turned
on around the various fixed points lies in the fact that the same method
is applicable to both normal as well as superconducting junctions of
any number of \qwd.

Now, for the case of three \ll wires meeting at a normal junction,
let us assume that the wires are parameterized by spatial
coordinates $x_i$ which go from zero to infinity with the junction being
situated at $x_i=0$ ($i$ being the wire index). The junction can be
parameterized by a $ 3\times 3$ \smat-matrix with diagonal elements
$r_{ii}$ and off-diagonal elements $t_{ij}$. In the presence of
$\ed$-$\ed$ interactions, the \rg equations can be
derived~\cite{yue,lal} by first expanding the electron wave-function
in each of the wires in terms of reflected and transmitted electron
waves (scattering wave basis). Then the amplitude of scattering of
electrons from the Friedel oscillations in the wires can be deduced
by using a Hartree-Fock decomposition of the $\ed$-$\ed$ interaction
term. Finally the \rg equation is obtained by applying the
poor-man's scaling approach~\cite{matveev}. The \rg equations for
the entire \smat-matrix can be written in a concise and compact from
given by~\cite{lal}
\beq \dfrac{d{\mathbb{S}}}{dl} = {\mathbb{F}} - \mathbb{S}
\mathbb{F}^{\dagger} \mathbb{S}~, \label{srg} \eeq
where, $l={\rm ln}(L/d)$ is the dimension-less \rg scale ($L$
corresponds to the physical length scale or energy scale at which we are
probing the system and $d$ is the short distance or  high energy
cut-off). Here ${\mathbb{F}}$ is a diagonal matrix with
${\mathbb{F}}_{ii} = -\alpha r_{ii}/2$ and $\alpha$ is the repulsive
$\ed$-$\ed$ interaction parameter which is related to the \ll
parameter $K$ as  $K=((1-\alpha)/(1+\alpha))^{1/2}$.

Analogously, the \nsn junction can be described  in terms of
an \smat-matrix with elements describing transmission of both electrons
and holes and their mixing at the junction. The corresponding \rg
equation for the \smat-matrix was obtained by the present
authors~\cite{nsnepl_drs,nsnprb_drs} which was an extension of
Eq.~\ref{srg} to the superconducting case. Here too for the \nsn
case, we will assume that the two wires are parameterized by spatial
coordinates $x_1$ and $x_2$ where
 $x_1,x_2$ vary from zero to infinity and that the junction is at
 $x_1=0=x_2$.
The presence of the superconductor is encoded in the parametrization of
the \smat-matrix representing the junction. Of course, this way of
accounting for the presence of the superconductor assumes that the
superconductor imposes static boundary conditions on the two wires
forming the \nsn junction. This is a valid
approximation as long as one is focusing on sub-gap transport
properties of the junction. We also assume
that the superconductor at the junction is a singlet superconductor;
hence the spin of the incident electron or hole is conserved as
it  scatters off the junction. This  results in a block diagonal form
of the \smat-matrix with each spin block being a $4\times 4 $ matrix
representing scattering of electrons and holes within the given
spin sector.

The \smat-matrix at the superconducting junction for the
spin-up, spin-down, electron-hole and left-right symmetric (symmetry
in wire index) case (suppressing the wire index)
 can be parameterized by
 $r$, the normal reflection amplitude, $r_{A}$, the \ar amplitude, $t_A$,
the \car
amplitude~\cite{hekking1,hekking2} and $t$, the transmission amplitude for both
electrons and holes. The
fermion fields can then be expanded around left and right Fermi points on each
wire as
$ \psi_{is}(x) = \Psi_{I\,is}(x)\,
e^{i\,k_F\,x} \,+\, \Psi_{O\,is}(x)\, e^{-i\,k_F\,x} ~; $
where $i$ is the wire index, $s$ is the spin index which can be
$\up,\dn$ and $I(O)$ stands for incoming (outgoing) fields. Note
that $\Psi_{I(O)}(x) $ are slowly varying fields on the scale of
$k_F^{-1}$.  Electrons with momenta $k$ in vicinity of $k_F$, on
each wire at position $x$ is given by
\bea
\Psi_{is}(x) &=& \int_{0}^{\infty} dk \Big[ b_{ks}^{} \epp +
d_{ks}^\dagger  \epm  \non\\
&+& r_{} ^{}  b_{ks}^{} \emp  +
  r^\star_{} d_{ks}^\dagger  \emm  \non\\
  &+& r_{}^{} d_{ks}^{} \emm + r^\star b_{ks}^\dagger \emp
\Big]
\label{equation1} \eea where $b_{ks}^{}$ is the particle destruction
operator and $d_{ks}^{}$ is the hole destruction operator and we
have allowed for non-conservation of charge due to the proximity
effect induced by the superconductor. We then allow for short-range
density-density interactions between the fermions,
\bea \hmi = \dfrac{1}{2} \, \int dx \,dy \,(\sum _{s} \rho_{is}^{})
V(x-y) \,(\sum_{s'}\rho_{is'}^{})~, \eea
Following the  procedure outlined in Ref.~\onlinecite{nsnprb_drs}, we
find that the \rg equation for the ${\mathbb{S}}$-matrix continues
to be of the form given in Eq.~\ref{srg}, but now ${\mathbb{F}}$ is
a non-diagonal matrix, \beq {\mathbb{F}} =
\begin{bmatrix}
~{\alpha r}/{2}& 0 & {-\alpha^\prime r_{A}}/{2} &0~  \\
~0& {\alpha r}/{2}& 0 & {-\alpha^\prime r_{A}}/{2} ~ \\
~{-\alpha^\prime r_{A}}/{2} & 0 & {\alpha r}/{2}& 0 ~\\
~0& {-\alpha^\prime r_{A}}/{2} & 0 & {\alpha r}/{2}~ \\
 \end{bmatrix}
\label{fmat} \eeq where $\alpha$ and $\alpha^{\prime}$ are the
$x$-independent part of the mean field amplitudes for Friedel
oscillations and the proximity induced pair potential inside the \qw
respectively. Generalization to particle-hole non-symmetric
situations will make this matrix asymmetric. It is worth pointing
out that even though the expression for electron field in
Eq.~\ref{equation1} assumes particle-hole symmetry which leads to
considerable simplification in the derivation for the \rg equation
(Eq.~\ref{srg}), our formalism is more general.
The  \rg equation (with appropriate
modification of the $\mathbb{F}$ matrix for the asymmetric case)
will also hold for \smat-matrices representing situations where the wire
index symmetry as well as the particle-hole symmetry is broken.

We will mainly focus on three different fixed points - the \cfp and \gfp
of a normal junction of three \ll wires~\cite{lal,meden} and the
symmetric fixed point (\sfpd)~\cite{nsnprb_drs} of the \nsn junction.
First we discuss the stability around the \cfp and \gfp of a normal
junction of three \ll wires (Y-junction) to benchmark our calculation with
known results~\cite{lal}. As a first step towards performing a systematic
stability analysis, we need to obtain an ${\mathbb{S}}$-matrix which
results from a very small unitary deviation from the fixed point
${\mathbb{S}}$-matrix. Given the number of independent parameters of
the ${\mathbb{S}}$-matrix dictated by symmetry and unitarity
constraints, the most general deviation from the fixed point
${\mathbb{S}}$-matrix can be obtained by multiplying the fixed point
${\mathbb{S}}$-matrix by another unitary matrix which is such that
it allows for a straightforward expansion in terms of small
parameters around the identity matrix. This is realized as follows -
\beq {\mathbb{S}} = {\mathbb{S}}_{0} \exp\left\{i \sum_{j=1}^{9}
\epsilon_{j}\lambda_{j}\right\}~, \label{smatn} \eeq
where $\mathbb{S}_0$ represents the fixed point
${\mathbb{S}}$-matrix and $\lambda_{j}$'s (along with the identity
$\lambda_0 =I$) are the eight generators of the $SU(3)$ group which
are traceless hermitian matrices. This can be straightforwardly
generalized to the case of N wires by using $SU(N)$ matrices.
Perturbations around these fixed points are characterized in terms
of the $\epsilon_{j}$'s. Of course, the resulting
$\mathbb{S}$-matrix obtained in this way  corresponds to a small
unitary deviation when $\epsilon_{j}$'s are small. To first order in
$\epsilon_{j}$'s, Eq.~\ref{smatn} reduces to
\beq {\mathbb{S}} = {\mathbb{S}}_{0}\left({\mathbb{I}} + i \sum_{j}
\epsilon_{j}\lambda_{j}\right)~, \label{smatd} \eeq
where ${\mathbb{S}}_{0}$ for  \cfp and \gfp fixed points are~\cite{lal} %
\beq
{\mathbb{S}}_{0}^{\cfp} = \begin{bmatrix} ~0 & 1 & 0 ~\\
~0 & 0 & 1 ~\\
~1 & 0 & 0 ~
\end{bmatrix}~; ~
{\mathbb{S}}_{0}^{\gfp} =
\begin{bmatrix} ~-1/3 & 2/3 & 2/3 ~\\
~2/3 & -1/3 & 2/3 ~\\
~2/3 & 2/3 & -1/3 ~
\end{bmatrix}~,
\label{sgfp} \eeq respectively. Using Eq.~\ref{smatd}, the \rg
equation (Eq.~\ref{srg}) with $\mathbb{S}$ expanded  to the linear
order in $\epsilon_j$ becomes
\bea i \sum_{j=1}^{9} \lambda_{j} \dfrac{d\epsilon_{j}}{dl} &=&
\mathbb{S}_{0}^{\dagger} \Big[
\mathbb{I}-i\sum_{j}\epsilon_{j}\lambda_{j} \Big] ~ \bigg\{
\mathbb{F}-\mathbb{S}_{0}
\Big[\mathbb{I}+i\sum_{j}\epsilon_{j}\lambda_{j}\Big] \non\\
&& \mathbb{F}^{\dagger} \Big[\mathbb{I}+i\sum_{j}
\epsilon_{j}\lambda_{j}\Big]\bigg\}~, \label{eqnm} \eea
where $\mathbb{F}$ is the diagonal part of the following quantity
\beq {\mathbb{F}}=\dfrac{\alpha}{2}~
\mathbb{S}_{0}\Big[\mathbb{I}+i\sum_{j}\epsilon_{j}\lambda_{j}\Big]_{\rm
{diagonal}}~. \label{fmatd} \eeq
 Restricting the RHS of Eq.~\ref{eqnm} to linear order in $\epsilon_j $'s,
 one then obtains
nine coupled  linear differential equations. Next by applying a
unitary rotation, we can decouple these coupled equations
(Eq.~\ref{eqnm}) and re-cast them in terms of new variables
$\epsilon_j^\prime$ (which are linear combinations of the original
$\epsilon_j$). The equations are now
 given by \beq \dfrac{d\epsilon_{j}^{\prime}} {dl} = \mu_j \,
\epsilon_{j}^{\prime} \eeq where $\mu_j$ is a real number
corresponding  to the `power
law' associated with perturbations turned on along each of the new
nine eigen-directions $\epsilon^\prime_j$. $\mu_j<0$ indicates
that the given direction is stable and $\mu_j>0$ indicates that
it is unstable.
 Here the non-diagonal $\epsilon_{j}$ are related to the diagonal
$\epsilon_{j}^{\prime}$ by  $\epsilon_{j}=\sum_{i} \mathbb{U}_{ji}
\, \epsilon_{i}^{\prime}$ where $\mathbb{U}$ is the diagonalizing
rotation matrix.

Hence we obtain all the power laws associated with the independent
perturbations that can be switched on around a given fixed point
\smat-matrix. Now it is straightforward to show that the power laws
associated with the
 \cfp and \gfp are given by [$\alpha/2$, $\alpha/2$, 0,
$\alpha/2$, $\alpha/2$, $\alpha/2$, $\alpha/2$, 0, 0] and [0, 0, 0,
0, 0, -$\alpha/3$, $2\alpha/3$, $2\alpha/3$, $\alpha$] respectively
which is consistent with results obtained in
Ref.~\onlinecite{lal}~\footnote{There is a correction to the  power
law obtained for the \gfp for the three-wire junction in
Ref.~\onlinecite{lal}. Lal et. al. had predicted a stable direction
with power law $-\alpha$ which should be corrected to $-
\alpha/3$~\cite{diptiman} as is obtained in this work. The corrected
power law is also consistent with that obtained in
Ref.~\onlinecite{volker} using a functional \rg procedure.}. The
value zero corresponds  to marginal directions while the values with
positive or negative signs correspond to stable or unstable
directions respectively. We do not write the explicit form of the
$\mathbb{U}$ matrix for the \cfp and \gfp as they are
needed only for obtaining the explicit form of the power law
correction to the  fixed point conductance which we do not calculate
for these cases.

Finally, let us discuss the stability around the different \rg fixed
points of the \nsn junction. First we focus on the \sfp of the \nsn
junction. In the presence of $\ed$-$\ed$ interaction
inside the \qw,  the incident
electron (hole) not only scatters from the Friedel oscillations as an
electron (hole) but also scatters from the proximity induced pair
potential inside the \qw as a hole (electron). Now the amplitude of both
of these scattering processes are proportional to the $\ed$-$\ed$
interaction strength inside the \qw. The competition between these two
scattering processes which actually arise due the same $\ed$-$\ed$
interaction strength inside the \qw leads to the presence of the new
\sfp where all the scattering amplitudes have intermediate non-zero
values. This fact is unique about this fixed point and hence this fixed
point is the central focus of our discussion. Details of this fixed
point are  further elaborated in the discussion at the end of this article.

We adopt the same procedure as described above for the three-wire junction
but now with $SU(4)$ generators. This is so because the full $8\times 8$
\smat-matrix describing the \nsn junction has a block diagonal form with
each spin block (up and down spin sectors) being represented by a $4\times
4$ matrix. Hence we have a unitary starting \smat-matrix deviating
from the fixed point \smat-matrix (\smat$_{0}$), as given before by
Eq.~\ref{smatd}, except that now the sum over $j$ runs  from $1$ to $16$
since $\lambda_{j}$'s now represent the fifteen generators of the
$SU(4)$ group along with the identity matrix. The \smat$_{0}$ which
describes the \sfpd ~\cite{nsnprb_drs} is given by $r=1/2$, $t=1/2$,
$r_{A}=-1/2$ and $t_{A}=1/2$. Note that the \sfp is a particle-hole,
left-right symmetric fixed point and hence the entire $4\times 4 $
\smat-matrix is determined completely by the above given four
amplitudes for $r,t,r_A,t_A$.

We then solve Eq.~\ref{eqnm} for this case with sixteen coupled
equations up to the first order in the small perturbations
$\epsilon_{j}$'s. We obtain the sixteen eigenvalues which correspond
to the power laws around the sixteen eigen-directions. These power
laws around the various eigen-directions can be listed as [0, 0, 0,
0, 0, $-\alpha/2$, $-\alpha/2$, $(\alpha-\alpha^{\prime})/2$,
$\alpha^{\prime}/2$, $\alpha^{\prime}/2$,
$(-\alpha+\alpha^{\prime})/2$, $(-\alpha+\alpha^{\prime})/2$,
$(\alpha+\alpha^{\prime})/2$, $(\alpha+\alpha^{\prime})/2$,
$(\alpha-\alpha^{\prime}-\sqrt{9\alpha^{2}+14\alpha
\alpha^{\prime}+9\alpha^{{\prime}^{2}}})/4$,
$(\alpha-\alpha^{\prime}+\sqrt{9\alpha^{2}+14\alpha
\alpha^{\prime}+9\alpha^{{\prime}^{2}}})/4$].\\
Hence we note that there are five marginal directions, two stable
directions, four unstable directions and four other directions whose
stability depends on the sign of $\alpha - \alpha^\prime$. One of
the most striking outcomes of this analysis is the fact that we
obtain two power laws which are not just simple linear combinations
of $V(0)$ and $V(2k_F)$. Instead,  they appear as square roots of quadratic
sum of these quantities. Our analysis actually leads to the first
demonstration of  the existence of such power laws  in the context of
quantum impurity problems in \ll theory and this is the central
result of this article.

Having obtained the power laws the next task is to obtain an
explicit expression for the Landauer-\butt conductance corresponding
to perturbations around these fixed points along some of
the eigen-directions. Now note that the \rg equation is expressed in
terms of $\epsilon^\prime$'s whereas the \smat-matrix representing
small deviations from the fixed point is expressed in terms of
$\epsilon$'s. The two terminal linear conductance across the
junction depends explicitly on the \smat-matrix element which are
expressed in terms of $\epsilon$'s (see Eq.~\ref{smatd}). Hence in
order to obtain an expression for conductance in terms of the
temperature or the applied voltage dependence induced by $\ed$-$\ed$
interaction, we need to first assign bare values to the various
perturbations parameterized by $\epsilon'$s and then  express the
$\epsilon'$'s evolved under \rg flow in terms of these bare values
of $\epsilon'$'s as $\epsilon'(\Lambda) = (\Lambda/\Lambda_0 )^\mu
\epsilon^\prime_0$ where $\Lambda$ corresponds to the energy scale
at which we are probing the system (which can be either voltage bias
at zero temperature or temperature at vanishing bias) and
$\Lambda_0$ is the high energy cut-off expressed in terms of voltage
or temperature. Then by using the rotation matrix which diagonalizes
the coupled \rg equations, we express $\epsilon$'s in terms of
$\epsilon'$'s written explicitly as a function of temperature or
voltage. Finally plugging these renormalized values of $\epsilon$'s
into the \smat-matrix given by Eq.~\ref{smatd}, we get all the
transmission and reflection amplitudes for the system as explicit
functions of the temperature or voltage carrying  the specific power laws
associated with perturbations switched on along the eigen-directions.
These amplitudes are now directly related to the linear conductances.\\
Now we will calculate expression for conductance for a simple case
where only one of the $\epsilon^{\prime}$($=\epsilon_{15}^{\prime}$)
is turned on. For this we need the $\mathbb{U}$ matrix for this case
which is given by
\begin{widetext}
\vskip-5mm
\bea \mathbb{U} &=& \left[ \begin{array}{cccccccccccccccc} ~0 & 0
& 0 & 0 & 0 & 0 & 0 & 0 & 0 & 0 & 0 & 0 & 0 & 0 & 0 & 1~
\\
~  0  &  0  &  \sqrt{{3}/{2}}  & 0 & 0  & 0  & 0  & -1/\sqrt{2} & 0
& 0  & 0  & 0 & 0
 & 0  & 1 & 0  ~
\\
~-1 & 0 & 1 & 0 & 0 & 0 & 0 & -\sqrt{3} & 0 & 0 & 0 & 0 & 1 & 0 & 0
& 0~
\\
~ 0 & 0 & 0 & -1 & 0 & 0 & 0 & 0 & 0 & 0 & 1 & 0 & 0 & 0 & 0 & 0 ~\\
~ 0 & 0 & 2 & 0 & 0 & -1 & 0 & 0 & 1 & 0 & 0 & 0 & 0 & 0 & 0 & 0 ~\\
~ 0 & -1 & 0 & 0 & -2 & 0 & -1 & 0 & 0 & -1 & 0 & 0 & 0 & 1 & 0 & 0 ~\\
~ 0 & 0 & 0 & 0 & 1 & 0 & 1 & 0 & 0 & 1 & 0 & 1 & 0 & 0 & 0 & 0 ~\\
~ 0 & 1 & 0 & 0 & 0 & 0 & -1 & 0 & 0 & 1 & 0 & 0 & 0 & 1 & 0 & 0 ~\\
~ 0 & 0 & -\sqrt{3/2} & 0 & 0 & -\sqrt{3/2} & 0 & -1/\sqrt{2} &
\sqrt{3/2} & 0 & 0 & 0 & 0 & 0 & 1 & 0 ~\\
~ -1 & 0 & 1 & 0 & 0 & 1 & 0 & \sqrt{3} & -1 & 0 & 0 & 0 & 1 & 0 & 0
&
0~\\
~ 0 & 1 & 0 & 0 & 0 & 0 & 1 & 0 & 0 & -1 & 0 & 0 & 0 & 1 & 0 & 0 ~\\
~ 1 & 0 & 0 & 0 & 0 & -1 & 0 & 0 & -1 & 0 & 0 & 0 & 1 & 0 & 0 & 0 ~\\
~ 0 & -1 & 0 & 0 & 2 & 0 & -1 & 0 & 0 & -1 & 0 & 0 & 0 & 1 & 0 & 0 ~\\
~ 0 & 0 & 0 & 0 & 1 & 0 & -1 & 0 & 0 & -1 & 0 & 1 & 0 & 0 & 0 & 0 ~\\
~ 1 & 0 & 0 & A & 0 & 1 & 0 & 0 & 1 & 0 & A & 0 & 1 & 0 & 0 & 0 ~\\
~ 1 & 0 & 0 & B & 0 & 1 & 0 & 0 & 1 & 0 & B & 0 & 1 & 0 & 0 & 0 ~\\
\end{array} \right]
\label{dumatrix} \eea
\end{widetext}
where, $A={4(\alpha + \alpha^{\prime})}/
{(\alpha-\alpha^{\prime}-\sqrt{9\alpha^{2}+14\alpha
\alpha^{\prime}+9\alpha^{{\prime}^{2}}})/4}$ and $B= {4(\alpha
+ \alpha^{\prime})}/
{(\alpha-\alpha^{\prime}+\sqrt{9\alpha^{2}+14\alpha
\alpha^{\prime}+9\alpha^{{\prime}^{2}}})/4}$.

We choose this specific direction to perturb  the system as this
corresponds to a power law which is not a linear function of $V(0)$
and $V(2K_F)$ and hence interesting to study. The \smat-matrix to
quadratic order in $\epsilon'_{15}$  is given by
\begin{widetext}
\vskip -0.55cm
\beq \mathbb{S} = \left[
\begin{array}{cccc} \frac{[1-(1-i)\epsilon_{15}^{\prime} - {\epsilon_{15}^{\prime}}^2]}{2} &
{\frac{[1+\epsilon_{15}^{\prime}]}{2}}-{\frac{[1-i]{\epsilon_{15}^{\prime}}^2}{4}}
& -\frac{1}{2} &
{\frac{[1+i\epsilon_{15}^{\prime}]}{2}}-{\frac{[1+i]{\epsilon_{15}^{\prime}
}^2}{4}}\\\\
\frac{[1-(1+i)\epsilon_{15}^{\prime} -
{\epsilon_{15}^{\prime}}^2]}{2} &
{\frac{[1+\epsilon_{15}^{\prime}]}{2}}-{\frac{[1+i]{\epsilon_{15}^{\prime}}^2}{4}}&
\frac{1}{2} &
{\frac{[1-i\epsilon_{15}^{\prime}]}{2}}+{\frac{[1-i]{\epsilon_{15}^{\prime}
}^2}{4}}\\\\
-\frac{[1+(1-i)\epsilon_{15}^{\prime} -
{\epsilon_{15}^{\prime}}^2]}{2} &
 {\frac{[1-\epsilon_{15}^{\prime}]}{2}}-{\frac{[1-i]{\epsilon_{15}^{\prime}}^2}{4}}
& \frac{1}{2} &
{\frac{[1-i\epsilon_{15}^{\prime}]}{2}}-{\frac{[1-i]{\epsilon_{15}^{\prime}
}^2}{4}} \\\\
\frac{[1+(1+i)\epsilon_{15}^{\prime} -
{\epsilon_{15}^{\prime}}^2]}{2} &
-{\frac{[1-\epsilon_{15}^{\prime}]}{2}}+{\frac{[1+i]{\epsilon_{15}^{\prime}}^2}{4}}
& \frac{1}{2} &
{\frac{[1+i\epsilon_{15}^{\prime}]}{2}}+{\frac{[1-i]{\epsilon_{15}^{\prime}}^2}{4}}
\end{array} \right]
\label{sdeviation}
\eeq
\end{widetext}
So, the scaling of sub-gap conductance (to ${\mathcal
O}(\epsilon_{15}^{\prime 2})$) for an incident electron and a hole
taking into account both spin-up and spin-down contributions in
units of $2e^2/h$ is given by
\bea
G_{12}^{e} = -\dfrac{\epsilon_{15}^{\prime}} {2}~; &&
G_{21}^{e} = \dfrac{\epsilon_{15}^{\prime}}{2} \eea
\bea G_{12}^{h} = -\dfrac{\epsilon_{15}^{\prime}}{2}~;  &&
G_{21}^{h} = \dfrac{\epsilon_{15}^{\prime}}{2}
\eea
where $\epsilon_{15}^{\prime}=\epsilon_{15,\,0}^{\prime}
 (\Lambda/\Lambda_0)^{(\alpha-\alpha^{\prime}-\sqrt{9\alpha^{2}+14\alpha
\alpha^{\prime}+9\alpha^{{\prime}^{2}}})/4}$. Here the superscripts $e$
and $h$ stand for electrons and holes while the subscripts $1$ and
$2$ stand for first and second  wire  respectively. Also,
$G_{12}^{e}=|t^{eh}_{A,\,12}|^{2} - |t^{ee}_{12}|^{2}$ where
$t^{ee}$ is the transmission amplitude for electrons and $t^{eh}_{A}$
represents \car amplitude for electrons. Similar expressions hold for the
holes. In the expressions of power laws given above,
$\alpha=(g_{2}-2g_{1})/2\pi \hbar v_{F}$ and
$\alpha^{\prime}=(g_{1}+g_{2})/2\pi \hbar v_{F}$ where the bare
values of $g_{1}(d)=V(2k_{F})$ and $g_{2}(d)=V(0)$. In our stability
analysis, we have assumed $\alpha < \alpha^{\prime}$ which is
consistent with experimental observations~\cite{yacoby2}. For the
special case when $g_{2}=2 g_{1}$, $\alpha$ vanishes and only
$\alpha^{\prime}$ survives.

It is very interesting to note that even though the
$\mathbb{S}$-matrix corresponding to perturbation along
$\epsilon^{\prime}_{15}$ breaks both time reversal and electron-hole
symmetry, the two terminal linear conductance restores particle-hole
symmetry. Secondly it might be of interest to note the fact that the
fixed point conductance admits correction along
$\epsilon^{\prime}_{15}$ which is linear in $\epsilon^{\prime}_{15}$
and not quadratic. Normally when we perform a stability analysis
around a fixed point $\mathbb{S}$-matrix whose elements are
constituted out of unimodular numbers (representing disconnected or
perfectly connected fixed points), it is  always possible to
identify  various terms of the $\mathbb{S}$-matrix, representing
small unitary deviations from  the fixed point
${\mathbb{S}}_0$-matrix in terms of various tunneling operators
which are perturbatively turned on around the fixed point
Hamiltonian. Hence a straight forward perturbative linear
conductance calculation using the Hamiltonian along with the
tunneling parts will suggest that the correction due to the
$\mathbb{S}$-matrix representing small deviation from  fixed point
${\mathbb{S}}_0$-matrix must introduce correction to fixed point
conductance which are quadratic in terms of the deviation parameter.
But this argument applies only to those fixed points which
correspond to completely connected or disconnected wires and not to
fixed points which have intermediate values for various transmission
and reflection amplitudes like the \sfp. In other words, an
arbitrary deviation from \sfp may not be easily representable as a
tunneling operator. This explains why the linear dependence of the
conductance on $\epsilon^{\prime}$ and hence the corresponding power
law dependence looks unconventional.

As a cross check,  we see that we get back the power laws associated
with the symmetric fixed point~\cite{lal} of the four-wire junction
once we substitute $\alpha^{\prime}=0$ in the expression for the
power laws of the \sfp for the \nsn junction. Although our geometry does
not correspond to the real junction of
four \ll wires, the presence of both electron and hole channel
mimics the situation of a four-wire junction. More specifically,
the symmetric fixed point of the \nsn junction (\sfpd) turns out to
be identical to the symmetric fixed point of the four-wire junction
due to perfect particle-hole symmetry of the \sfp when
$\alpha^{\prime}$ is set to zero.

Next we enumerate and discuss the
stability of the other fixed points (\rfpd, \afpd, \tfp and \cafpd)
obtained in Ref.~\onlinecite{nsnprb_drs} for the  \nsn junction :
\begin{enumerate}
\item[] (a) $t=t_{A}=r_A=0, r=1$  (\rfpd) :  This fixed point
turns out to be stable against perturbations in all directions.
There are ten directions for which the exponent  is -$\alpha$ while
two others  with the exponents -($\alpha$+$\alpha^{\prime}$). The
remaining four directions are marginal.
\item[] (b) $t=t_{A}=r=0, r_A=1$ (\afpd) : This is unstable against
perturbations in twelve directions. There are ten directions with
exponent $\alpha$ and two directions with exponent
($\alpha$+$\alpha^{\prime}$). The remaining four directions are
marginal,  as for \rfpd.
\item[] (c)  $r_{A}=t_{A}=r=0, t=1$ (\tfpd) : This fixed point has
four unstable directions with exponent  $\alpha$, two stable
directions with the exponent  -$\alpha^{\prime}$ and  the remaining
directions are marginal.
\item[] (d) $r_{A}=t=r=0, t_A=1$ (\cafpd) : This  has four unstable
directions with exponent $\alpha$ and two stable directions with the
exponent  -$\alpha^{\prime}$ and the remaining directions are marginal.
\end{enumerate}
Note the close similarity in stability between \cafp and \tfp fixed
points. This can be attributed to the fact that both these fixed
points belong to the continuous family of marginal fixed points
defined by the condition $|t|^2 + |t_{A}|^2=1$. The entire family of
fixed points is marginal because for these fixed points, the amplitudes
for Friedel oscillation and pair potential in the wire vanish identically.

Hence,  we notice that for the \afp only the scattering amplitude from the
pair potential inside the \qw is non-zero as only $r_A$ is nonzero,
and  for \rfp  only the scattering  amplitude from Friedel
oscillations are non-zero as only $r$ is nonzero. Furthermore,  both
for \cafp  and \tfp, the amplitude for scattering from the
Friedel oscillations as well as from the pair potential is zero as in these
cases both $r$ and $r_A$ are zero. So \sfp is the only fixed point for
which both the amplitude for scattering from the Friedel oscillations and
the pair amplitude are finite; hence, this fixed point is nontrivial. Its
very existence can be attributed to the interplay of these two different
scattering processes arising from Friedel oscillations and the  pair
potential inside the wire. The conductance at this fixed point gets
contribution from both the elastic co-tunneling (\ctd) of electrons through
the superconductor as well as through the crossed Andreev reflection
(\card) process. Since both electron and hole channels contribute with
opposite signs to conductance, if we give a small perturbation around this
fixed point, we get an interesting non-monotonic behavior of the
conductance $G_{\nsn}=G_{\car}~-~G_{\ct}$. This effect emerges due to the
competition between the electron and the hole channel and it can be of
interest from an experimental point of view. Also note that at the \sfp,
\ct amplitude of electrons $t=1/2$ and the \car amplitude $t_{A}=1/2$. 
This means that if we have an incident spin-polarized beam of (say "up"
polarized) electrons on the junction, when the junction is tuned to this
fixed point, $25\%$ of the spin up electrons get transmitted through the
junction and $25\%$ of the spin up electrons get converted to spin up holes
as they pass through the junction. Hence the transmitted charge across the
junction is zero on the average, but there is pure spin current flowing out
of the junction. Equivalently, we can think that the pure spin current is
generated due to flow of two beams of electrons of equal intensity, one
with spin up electrons and the other with spin down electrons propagating
in opposite directions. Therefore the \sfp can be relevant for future
spintronics applications. These points have been discussed in detail in
Ref.~\onlinecite{nsnepl_drs}.

To summarize, we have laid down a scheme to perform a systematic
stability analysis which works well for both  normal and
superconducting junctions of multiple \ll \qwd. Using our procedure,
we reproduced the known power laws for the fixed points of the
three- and four-wire junctions. Then we applied it to the  \nsn
junction and established the existence of non-trivial power laws
which are non-linear functions of $V(0)$ and $V(2 k_F)$. Finally, we
calculated the Landauer-\butt conductance associated with the
perturbations switched on around these fixed points and found the
explicit voltage or temperature power law dependence.
\vspace{0.5cm}
%
\acknowledgements It is a pleasure to thank Diptiman Sen for useful
discussions. SD acknowledges financial support under the DST project
(SR/S2/CMP-27/2006). AS acknowledges the Centre for High Energy
Physics, Indian Institute of Science, Bangalore (India)  for warm
hospitality during the initial stages of this work.
\bibliographystyle{apsrev}
\bibliography{reflist}

\begin{thebibliography}{40}
\expandafter\ifx\csname natexlab\endcsname\relax\def\natexlab#1{#1}\fi
\expandafter\ifx\csname bibnamefont\endcsname\relax
  \def\bibnamefont#1{#1}\fi
\expandafter\ifx\csname bibfnamefont\endcsname\relax
  \def\bibfnamefont#1{#1}\fi
\expandafter\ifx\csname citenamefont\endcsname\relax
  \def\citenamefont#1{#1}\fi
\expandafter\ifx\csname url\endcsname\relax
  \def\url#1{\texttt{#1}}\fi
\expandafter\ifx\csname urlprefix\endcsname\relax\def\urlprefix{URL }\fi
\providecommand{\bibinfo}[2]{#2}
\providecommand{\eprint}[2][]{\url{#2}}

\bibitem[{\citenamefont{Kane and Fisher}(1992)}]{k&f}
\bibinfo{author}{\bibfnamefont{C.~L.} \bibnamefont{Kane}} \bibnamefont{and}
  \bibinfo{author}{\bibfnamefont{M.~P.~A.} \bibnamefont{Fisher}},
  \bibinfo{journal}{Phys. Rev. B} \textbf{\bibinfo{volume}{46}},
  \bibinfo{pages}{15233} (\bibinfo{year}{1992}).

\bibitem[{\citenamefont{Yue et~al.}(1994)\citenamefont{Yue, Glazman, and
  Matveev}}]{yue}
\bibinfo{author}{\bibfnamefont{D.}~\bibnamefont{Yue}},
  \bibinfo{author}{\bibfnamefont{L.~I.} \bibnamefont{Glazman}},
  \bibnamefont{and} \bibinfo{author}{\bibfnamefont{K.~A.}
  \bibnamefont{Matveev}}, \bibinfo{journal}{Phys. Rev. B}
  \textbf{\bibinfo{volume}{49}}, \bibinfo{pages}{1966} (\bibinfo{year}{1994}).

\bibitem[{\citenamefont{Nayak et~al.}(1999)\citenamefont{Nayak, Fisher, Ludwig,
  and Lin}}]{nayak}
\bibinfo{author}{\bibfnamefont{C.}~\bibnamefont{Nayak}},
  \bibinfo{author}{\bibfnamefont{M.~P.~A.} \bibnamefont{Fisher}},
  \bibinfo{author}{\bibfnamefont{A.~W.~W.} \bibnamefont{Ludwig}},
  \bibnamefont{and} \bibinfo{author}{\bibfnamefont{H.~H.} \bibnamefont{Lin}},
  \bibinfo{journal}{Phys. Rev. B} \textbf{\bibinfo{volume}{59}},
  \bibinfo{pages}{15694} (\bibinfo{year}{1999}).

\bibitem[{\citenamefont{Lal et~al.}(2002)\citenamefont{Lal, Rao, and
  Sen}}]{lal}
\bibinfo{author}{\bibfnamefont{S.}~\bibnamefont{Lal}},
  \bibinfo{author}{\bibfnamefont{S.}~\bibnamefont{Rao}}, \bibnamefont{and}
  \bibinfo{author}{\bibfnamefont{D.}~\bibnamefont{Sen}},
  \bibinfo{journal}{Phys. Rev. B} \textbf{\bibinfo{volume}{66}},
  \bibinfo{pages}{165327} (\bibinfo{year}{2002}).

\bibitem[{\citenamefont{Das et~al.}(2004)\citenamefont{Das, Rao, and
  Sen}}]{rgstudy_drs}
\bibinfo{author}{\bibfnamefont{S.}~\bibnamefont{Das}},
  \bibinfo{author}{\bibfnamefont{S.}~\bibnamefont{Rao}}, \bibnamefont{and}
  \bibinfo{author}{\bibfnamefont{D.}~\bibnamefont{Sen}},
  \bibinfo{journal}{Phys. Rev. B} \textbf{\bibinfo{volume}{70}},
  \bibinfo{eid}{085318} (\bibinfo{year}{2004}).

\bibitem[{\citenamefont{Chamon et~al.}(2003)\citenamefont{Chamon, Oshikawa, and
  Affleck}}]{affleck1}
\bibinfo{author}{\bibfnamefont{C.}~\bibnamefont{Chamon}},
  \bibinfo{author}{\bibfnamefont{M.}~\bibnamefont{Oshikawa}}, \bibnamefont{and}
  \bibinfo{author}{\bibfnamefont{I.}~\bibnamefont{Affleck}},
  \bibinfo{journal}{Phys. Rev. Lett.} \textbf{\bibinfo{volume}{91}},
  \bibinfo{pages}{206403} (\bibinfo{year}{2003}).

\bibitem[{\citenamefont{Oshikawa et~al.}(2006)\citenamefont{Oshikawa, Chamon,
  and Affleck}}]{affleck2}
\bibinfo{author}{\bibfnamefont{M.}~\bibnamefont{Oshikawa}},
  \bibinfo{author}{\bibfnamefont{C.}~\bibnamefont{Chamon}}, \bibnamefont{and}
  \bibinfo{author}{\bibfnamefont{I.}~\bibnamefont{Affleck}},
  \bibinfo{journal}{J. Stat. Mech.: Theory and Exp.}
  \textbf{\bibinfo{volume}{2006}}, \bibinfo{pages}{P02008}
  (\bibinfo{year}{2006}).

\bibitem[{\citenamefont{Bellazzini et~al.}(2007)\citenamefont{Bellazzini,
  Mintchev, and Sorba}}]{bellazzini2006}
\bibinfo{author}{\bibfnamefont{B.}~\bibnamefont{Bellazzini}},
  \bibinfo{author}{\bibfnamefont{M.}~\bibnamefont{Mintchev}}, \bibnamefont{and}
  \bibinfo{author}{\bibfnamefont{P.}~\bibnamefont{Sorba}},
  \bibinfo{journal}{J.Phys.A} \textbf{\bibinfo{volume}{40}},
  \bibinfo{pages}{2485} (\bibinfo{year}{2007}).

\bibitem[{\citenamefont{Bellazzini
  et~al.}(2008{\natexlab{a}})\citenamefont{Bellazzini, Burrello, Mintchev, and
  Sorba}}]{bellazzini12008}
\bibinfo{author}{\bibfnamefont{B.}~\bibnamefont{Bellazzini}},
  \bibinfo{author}{\bibfnamefont{M.}~\bibnamefont{Burrello}},
  \bibinfo{author}{\bibfnamefont{M.}~\bibnamefont{Mintchev}}, \bibnamefont{and}
  \bibinfo{author}{\bibfnamefont{P.}~\bibnamefont{Sorba}}
  (\bibinfo{year}{2008}{\natexlab{a}}), \bibinfo{note}{{{arXiv:0801.2852
  [hep-th]}}}.

\bibitem[{\citenamefont{Bellazzini
  et~al.}(2008{\natexlab{b}})\citenamefont{Bellazzini, Mintchev, and
  Sorba}}]{bellazzini22008}
\bibinfo{author}{\bibfnamefont{B.}~\bibnamefont{Bellazzini}},
  \bibinfo{author}{\bibfnamefont{M.}~\bibnamefont{Mintchev}}, \bibnamefont{and}
  \bibinfo{author}{\bibfnamefont{P.}~\bibnamefont{Sorba}}
  (\bibinfo{year}{2008}{\natexlab{b}}), \bibinfo{note}{{{arXiv:0810.3101
  [hep-th]}}}.

\bibitem[{\citenamefont{Agarwal et~al.}(2008)\citenamefont{Agarwal, Das, Rao,
  and Sen}}]{tdos_adrs}
\bibinfo{author}{\bibfnamefont{A.}~\bibnamefont{Agarwal}},
  \bibinfo{author}{\bibfnamefont{S.}~\bibnamefont{Das}},
  \bibinfo{author}{\bibfnamefont{S.}~\bibnamefont{Rao}}, \bibnamefont{and}
  \bibinfo{author}{\bibfnamefont{D.}~\bibnamefont{Sen}} (\bibinfo{year}{2008}),
  \eprint{arXiv/0810.3513}.

\bibitem[{\citenamefont{Takane and Koyama}(1996)}]{jap1}
\bibinfo{author}{\bibfnamefont{T.}~\bibnamefont{Takane}} \bibnamefont{and}
  \bibinfo{author}{\bibfnamefont{Y.}~\bibnamefont{Koyama}},
  \bibinfo{journal}{J. Phys. Soc. Jpn.} \textbf{\bibinfo{volume}{65}},
  \bibinfo{pages}{3630} (\bibinfo{year}{1996}).

\bibitem[{\citenamefont{Takane and Koyama}(1997)}]{jap3}
\bibinfo{author}{\bibfnamefont{T.}~\bibnamefont{Takane}} \bibnamefont{and}
  \bibinfo{author}{\bibfnamefont{Y.}~\bibnamefont{Koyama}},
  \bibinfo{journal}{J. Phys. Soc. Jpn.} \textbf{\bibinfo{volume}{66}},
  \bibinfo{pages}{419} (\bibinfo{year}{1997}).

\bibitem[{\citenamefont{Maslov et~al.}(1996)\citenamefont{Maslov, Stone,
  Golbert, and Loss}}]{maslov1996PRB}
\bibinfo{author}{\bibfnamefont{D.~L.} \bibnamefont{Maslov}},
  \bibinfo{author}{\bibfnamefont{M.}~\bibnamefont{Stone}},
  \bibinfo{author}{\bibfnamefont{P.~M.} \bibnamefont{Golbert}},
  \bibnamefont{and} \bibinfo{author}{\bibfnamefont{D.}~\bibnamefont{Loss}},
  \bibinfo{journal}{Phys. Rev. B} \textbf{\bibinfo{volume}{53}},
  \bibinfo{pages}{1548} (\bibinfo{year}{1996}).

\bibitem[{\citenamefont{{Fazio} et~al.}(1999)\citenamefont{{Fazio}, {Hekking},
  {Odintsov}, and {Raimondi}}}]{fazio}
\bibinfo{author}{\bibfnamefont{R.}~\bibnamefont{{Fazio}}},
  \bibinfo{author}{\bibfnamefont{F.~W.~J.} \bibnamefont{{Hekking}}},
  \bibinfo{author}{\bibfnamefont{A.~A.} \bibnamefont{{Odintsov}}},
  \bibnamefont{and}
  \bibinfo{author}{\bibfnamefont{R.}~\bibnamefont{{Raimondi}}},
  \bibinfo{journal}{Superlattices Microstruct.} \textbf{\bibinfo{volume}{25}},
  \bibinfo{pages}{1163} (\bibinfo{year}{1999}).

\bibitem[{\citenamefont{Vishveshwara et~al.}(2002)\citenamefont{Vishveshwara,
  Bena, Balents, and Fisher}}]{bena}
\bibinfo{author}{\bibfnamefont{S.}~\bibnamefont{Vishveshwara}},
  \bibinfo{author}{\bibfnamefont{C.}~\bibnamefont{Bena}},
  \bibinfo{author}{\bibfnamefont{L.}~\bibnamefont{Balents}}, \bibnamefont{and}
  \bibinfo{author}{\bibfnamefont{M.~P.~A.} \bibnamefont{Fisher}},
  \bibinfo{journal}{Phys. Rev. B} \textbf{\bibinfo{volume}{66}},
  \bibinfo{pages}{165411} (\bibinfo{year}{2002}).

\bibitem[{\citenamefont{Man et~al.}(2005)\citenamefont{Man, Klapwijk, and
  Morpurgo}}]{man}
\bibinfo{author}{\bibfnamefont{H.~T.} \bibnamefont{Man}},
  \bibinfo{author}{\bibfnamefont{T.~M.} \bibnamefont{Klapwijk}},
  \bibnamefont{and} \bibinfo{author}{\bibfnamefont{A.~F.}
  \bibnamefont{Morpurgo}} (\bibinfo{year}{2005}),
  \bibinfo{note}{{arXiv:cond-mat/0504566}}.

\bibitem[{\citenamefont{Recher and Loss}(2002)}]{dloss}
\bibinfo{author}{\bibfnamefont{P.}~\bibnamefont{Recher}} \bibnamefont{and}
  \bibinfo{author}{\bibfnamefont{D.}~\bibnamefont{Loss}},
  \bibinfo{journal}{Phys. Rev. B} \textbf{\bibinfo{volume}{65}},
  \bibinfo{pages}{165327} (\bibinfo{year}{2002}).

\bibitem[{\citenamefont{Titov et~al.}(2006)\citenamefont{Titov, M\"{u}ller, and
  Belzig}}]{titov}
\bibinfo{author}{\bibfnamefont{M.}~\bibnamefont{Titov}},
  \bibinfo{author}{\bibfnamefont{M.}~\bibnamefont{M\"{u}ller}},
  \bibnamefont{and} \bibinfo{author}{\bibfnamefont{W.}~\bibnamefont{Belzig}},
  \bibinfo{journal}{Phys. Rev. Lett.} \textbf{\bibinfo{volume}{97}},
  \bibinfo{eid}{237006} (\bibinfo{year}{2006}).

\bibitem[{\citenamefont{Winkelholz et~al.}(1996)\citenamefont{Winkelholz,
  Fazio, Hekking, and Sch\"{o}n}}]{winkelholz}
\bibinfo{author}{\bibfnamefont{C.}~\bibnamefont{Winkelholz}},
  \bibinfo{author}{\bibfnamefont{R.}~\bibnamefont{Fazio}},
  \bibinfo{author}{\bibfnamefont{F.~W.~J.} \bibnamefont{Hekking}},
  \bibnamefont{and}
  \bibinfo{author}{\bibfnamefont{G.}~\bibnamefont{Sch\"{o}n}},
  \bibinfo{journal}{Phys. Rev. Lett.} \textbf{\bibinfo{volume}{77}},
  \bibinfo{pages}{3200} (\bibinfo{year}{1996}).

\bibitem[{\citenamefont{Hou and Chamon}(2008)}]{chamon2008}
\bibinfo{author}{\bibfnamefont{C.~Y.} \bibnamefont{Hou}} \bibnamefont{and}
  \bibinfo{author}{\bibfnamefont{C.}~\bibnamefont{Chamon}},
  \bibinfo{journal}{Phys. Rev. B} \textbf{\bibinfo{volume}{77}},
  \bibinfo{pages}{155422} (\bibinfo{year}{2008}).

\bibitem[{\citenamefont{Das et~al.}(2008{\natexlab{a}})\citenamefont{Das, Rao,
  and Saha}}]{nsnepl_drs}
\bibinfo{author}{\bibfnamefont{S.}~\bibnamefont{Das}},
  \bibinfo{author}{\bibfnamefont{S.}~\bibnamefont{Rao}}, \bibnamefont{and}
  \bibinfo{author}{\bibfnamefont{A.}~\bibnamefont{Saha}},
  \bibinfo{journal}{Europhys. Lett.} \textbf{\bibinfo{volume}{81}},
  \bibinfo{pages}{67001} (\bibinfo{year}{2008}{\natexlab{a}}).

\bibitem[{\citenamefont{Das et~al.}(2008{\natexlab{b}})\citenamefont{Das, Rao,
  and Saha}}]{nsnprb_drs}
\bibinfo{author}{\bibfnamefont{S.}~\bibnamefont{Das}},
  \bibinfo{author}{\bibfnamefont{S.}~\bibnamefont{Rao}}, \bibnamefont{and}
  \bibinfo{author}{\bibfnamefont{A.}~\bibnamefont{Saha}},
  \bibinfo{journal}{Phys. Rev. B} \textbf{\bibinfo{volume}{77}},
  \bibinfo{eid}{155418} (\bibinfo{year}{2008}{\natexlab{b}}).

\bibitem[{\citenamefont{Das and Rao}(2008)}]{superduality_dr}
\bibinfo{author}{\bibfnamefont{S.}~\bibnamefont{Das}} \bibnamefont{and}
  \bibinfo{author}{\bibfnamefont{S.}~\bibnamefont{Rao}},
  \bibinfo{journal}{Phys. Rev. B} \textbf{\bibinfo{volume}{78}},
  \bibinfo{eid}{205421} (\bibinfo{year}{2008}).

\bibitem[{\citenamefont{{Falci} et~al.}(2001)\citenamefont{{Falci}, {Feinberg},
  and {Hekking}}}]{hekking1}
\bibinfo{author}{\bibfnamefont{G.}~\bibnamefont{{Falci}}},
  \bibinfo{author}{\bibfnamefont{D.}~\bibnamefont{{Feinberg}}},
  \bibnamefont{and} \bibinfo{author}{\bibfnamefont{F.~W.~J.}
  \bibnamefont{{Hekking}}}, \bibinfo{journal}{Europhys. Lett.}
  \textbf{\bibinfo{volume}{54}}, \bibinfo{pages}{255} (\bibinfo{year}{2001}).

\bibitem[{\citenamefont{{Bignon} et~al.}(2004)\citenamefont{{Bignon}, {Houzet},
  {Pistolesi}, and {Hekking}}}]{hekking2}
\bibinfo{author}{\bibfnamefont{G.}~\bibnamefont{{Bignon}}},
  \bibinfo{author}{\bibfnamefont{M.}~\bibnamefont{{Houzet}}},
  \bibinfo{author}{\bibfnamefont{F.}~\bibnamefont{{Pistolesi}}},
  \bibnamefont{and} \bibinfo{author}{\bibfnamefont{F.~W.~J.}
  \bibnamefont{{Hekking}}}, \bibinfo{journal}{Europhys. Lett.}
  \textbf{\bibinfo{volume}{67}}, \bibinfo{pages}{110} (\bibinfo{year}{2004}).

\bibitem[{\citenamefont{Russo et~al.}(2005)\citenamefont{Russo, Kroug,
  Klapwijk, and Morpurgo}}]{russo}
\bibinfo{author}{\bibfnamefont{S.}~\bibnamefont{Russo}},
  \bibinfo{author}{\bibfnamefont{M.}~\bibnamefont{Kroug}},
  \bibinfo{author}{\bibfnamefont{T.~M.} \bibnamefont{Klapwijk}},
  \bibnamefont{and} \bibinfo{author}{\bibfnamefont{A.~F.}
  \bibnamefont{Morpurgo}}, \bibinfo{journal}{Phys. Rev. Lett.}
  \textbf{\bibinfo{volume}{95}}, \bibinfo{eid}{027002} (\bibinfo{year}{2005}).

\bibitem[{\citenamefont{Cadden-Zimansky and
  Chandrasekhar}(2006)}]{chandrasekhar}
\bibinfo{author}{\bibfnamefont{P.}~\bibnamefont{Cadden-Zimansky}}
  \bibnamefont{and}
  \bibinfo{author}{\bibfnamefont{V.}~\bibnamefont{Chandrasekhar}},
  \bibinfo{journal}{Phys. Rev. Lett.} \textbf{\bibinfo{volume}{97}},
  \bibinfo{eid}{237003} (\bibinfo{year}{2006}).

\bibitem[{\citenamefont{Yeyati et~al.}(2007)\citenamefont{Yeyati, Bergeret,
  Martin-Rodero, and Klapwijk}}]{yeyati}
\bibinfo{author}{\bibfnamefont{A.~L.} \bibnamefont{Yeyati}},
  \bibinfo{author}{\bibfnamefont{F.~S.} \bibnamefont{Bergeret}},
  \bibinfo{author}{\bibfnamefont{A.}~\bibnamefont{Martin-Rodero}},
  \bibnamefont{and} \bibinfo{author}{\bibfnamefont{T.~M.}
  \bibnamefont{Klapwijk}}, \bibinfo{journal}{Nat. Phys.}
  \textbf{\bibinfo{volume}{3}}, \bibinfo{pages}{455} (\bibinfo{year}{2007}).

\bibitem[{\citenamefont{Golubev and Zaikin}(2007)}]{zaikin2007}
\bibinfo{author}{\bibfnamefont{D.~S.} \bibnamefont{Golubev}} \bibnamefont{and}
  \bibinfo{author}{\bibfnamefont{A.~D.} \bibnamefont{Zaikin}},
  \bibinfo{journal}{Phys. Rev. B} \textbf{\bibinfo{volume}{76}},
  \bibinfo{pages}{184510} (\bibinfo{year}{2007}).

\bibitem[{\citenamefont{Kalenkov and Zaikin}(2007)}]{zaikin12007}
\bibinfo{author}{\bibfnamefont{M.~S.} \bibnamefont{Kalenkov}} \bibnamefont{and}
  \bibinfo{author}{\bibfnamefont{A.~D.} \bibnamefont{Zaikin}},
  \bibinfo{journal}{Phys. Rev. B} \textbf{\bibinfo{volume}{76}},
  \bibinfo{pages}{224506} (\bibinfo{year}{2007}).

\bibitem[{\citenamefont{Futterer et~al.}(2009)\citenamefont{Futterer,
  Governale, Pala, and K\"{o}nig}}]{konig2009}
\bibinfo{author}{\bibfnamefont{D.}~\bibnamefont{Futterer}},
  \bibinfo{author}{\bibfnamefont{M.}~\bibnamefont{Governale}},
  \bibinfo{author}{\bibfnamefont{M.~G.} \bibnamefont{Pala}}, \bibnamefont{and}
  \bibinfo{author}{\bibfnamefont{J.}~\bibnamefont{K\"{o}nig}},
  \bibinfo{journal}{Phys. Rev. B} \textbf{\bibinfo{volume}{79}},
  \bibinfo{pages}{054505} (\bibinfo{year}{2009}).

\bibitem[{\citenamefont{Morten et~al.}(2008)\citenamefont{Morten,
  Huertas-Hernando, Belzig, and Brataas}}]{belzig2008}
\bibinfo{author}{\bibfnamefont{J.~P.} \bibnamefont{Morten}},
  \bibinfo{author}{\bibfnamefont{D.}~\bibnamefont{Huertas-Hernando}},
  \bibinfo{author}{\bibfnamefont{W.}~\bibnamefont{Belzig}}, \bibnamefont{and}
  \bibinfo{author}{\bibfnamefont{A.}~\bibnamefont{Brataas}},
  \bibinfo{journal}{Phys. Rev. B} \textbf{\bibinfo{volume}{78}},
  \bibinfo{pages}{224515} (\bibinfo{year}{2008}).

\bibitem[{\citenamefont{Morten et~al.}(2006)\citenamefont{Morten, Brataas, and
  Belzig}}]{belzig2006}
\bibinfo{author}{\bibfnamefont{J.~P.} \bibnamefont{Morten}},
  \bibinfo{author}{\bibfnamefont{A.}~\bibnamefont{Brataas}}, \bibnamefont{and}
  \bibinfo{author}{\bibfnamefont{W.}~\bibnamefont{Belzig}},
  \bibinfo{journal}{Phys. Rev. B} \textbf{\bibinfo{volume}{74}},
  \bibinfo{pages}{214510} (\bibinfo{year}{2006}).

\bibitem[{\citenamefont{Golubev and Zaikin}(2009)}]{zaikin2009}
\bibinfo{author}{\bibfnamefont{D.}~\bibnamefont{Golubev}} \bibnamefont{and}
  \bibinfo{author}{\bibfnamefont{A.}~\bibnamefont{Zaikin}}
  (\bibinfo{year}{2009}), \bibinfo{note}{{{arXiv:0902.2864 [cond-mat]}}}.

\bibitem[{\citenamefont{Matveev et~al.}(1993)\citenamefont{Matveev, Yue, and
  Glazman}}]{matveev}
\bibinfo{author}{\bibfnamefont{K.~A.} \bibnamefont{Matveev}},
  \bibinfo{author}{\bibfnamefont{D.}~\bibnamefont{Yue}}, \bibnamefont{and}
  \bibinfo{author}{\bibfnamefont{L.~I.} \bibnamefont{Glazman}},
  \bibinfo{journal}{Phys. Rev. Lett.} \textbf{\bibinfo{volume}{71}},
  \bibinfo{pages}{3351} (\bibinfo{year}{1993}).

\bibitem[{\citenamefont{Barnabe-Theriault
  et~al.}(2005)\citenamefont{Barnabe-Theriault, Sedeki, Meden, and
  Scho¨nhammer}}]{meden}
\bibinfo{author}{\bibfnamefont{X.}~\bibnamefont{Barnabe-Theriault}},
  \bibinfo{author}{\bibfnamefont{A.}~\bibnamefont{Sedeki}},
  \bibinfo{author}{\bibfnamefont{V.}~\bibnamefont{Meden}}, \bibnamefont{and}
  \bibinfo{author}{\bibfnamefont{K.}~\bibnamefont{Scho¨nhammer}},
  \bibinfo{journal}{Phys. Rev. Lett} \textbf{\bibinfo{volume}{94}},
  \bibinfo{pages}{136405} (\bibinfo{year}{2005}).

\bibitem[{\citenamefont{Auslaender et~al.}(2002)\citenamefont{Auslaender,
  Yacoby, de~Picciotto, Baldwin, Pfeiffer, and West}}]{yacoby2}
\bibinfo{author}{\bibfnamefont{O.~M.} \bibnamefont{Auslaender}},
  \bibinfo{author}{\bibfnamefont{A.}~\bibnamefont{Yacoby}},
  \bibinfo{author}{\bibfnamefont{R.}~\bibnamefont{de~Picciotto}},
  \bibinfo{author}{\bibfnamefont{K.~W.} \bibnamefont{Baldwin}},
  \bibinfo{author}{\bibfnamefont{L.~N.} \bibnamefont{Pfeiffer}},
  \bibnamefont{and} \bibinfo{author}{\bibfnamefont{K.~W.} \bibnamefont{West}},
  \bibinfo{journal}{Science} \textbf{\bibinfo{volume}{295}},
  \bibinfo{pages}{825} (\bibinfo{year}{2002}).

\bibitem[{\citenamefont{Sen}()}]{diptiman}
\bibinfo{author}{\bibfnamefont{D.}~\bibnamefont{Sen}}, \bibinfo{note}{(private
  communication)}.

\bibitem[{\citenamefont{Meden et~al.}(2008)\citenamefont{Meden, Andergassen,
  Enss, Schoeller, and Schoenhammer}}]{volker}
\bibinfo{author}{\bibfnamefont{V.}~\bibnamefont{Meden}},
  \bibinfo{author}{\bibfnamefont{S.}~\bibnamefont{Andergassen}},
  \bibinfo{author}{\bibfnamefont{T.}~\bibnamefont{Enss}},
  \bibinfo{author}{\bibfnamefont{H.}~\bibnamefont{Schoeller}},
  \bibnamefont{and}
  \bibinfo{author}{\bibfnamefont{K.}~\bibnamefont{Schoenhammer}},
  \bibinfo{journal}{New Journal of Physics} \textbf{\bibinfo{volume}{10}},
  \bibinfo{pages}{045012} (\bibinfo{year}{2008}).

\end{thebibliography}
\end{document}